\documentclass[5p,twocolumn,preprint]{elsarticle}

\usepackage{amsmath}
\usepackage{amssymb}
\usepackage{graphicx}
\usepackage{upgreek}

\newcommand{\be}{\begin{equation}}
\newcommand{\ee}{\end{equation}}
\newcommand{\bs}{\begin{subequations}}
\newcommand{\es}{\end{subequations}}

\newcommand{\vt}{V_\theta}
\newcommand{\Ii}{I_1}
\newcommand{\Ki}{K_1}

\newcommand{\kr}{k_n r}
\newcommand{\ka}{k_n a}
\newcommand{\kb}{k_n b}
\newcommand{\kz}{k_n z}

\newcommand{\rmd}{\mathrm{d}}
\newcommand{\rmi}{\mathrm{i}}
\newcommand{\sg}{\mathrm{Sg}}

\newcommand{\pr}{\partial_r}
\newcommand{\pz}{\partial_z}
\newcommand{\half}{{\textstyle\frac1{2}}}

\begin{document}

\begin{frontmatter}

\title{Mixing layer between two co-current Taylor-Couette flows}

\author[ntnu]{Simen {\AA}. Ellingsen\corref{cor}\fnref{fax}}
\ead{simen.a.ellingsen@ntnu.no}
\author[ntnu]{Helge I. Andersson}

\cortext[cor]{Corresponding author.}
\fntext[fax]{Fax number: +47 73593491. Telephone: +47 73593554.}
\address[ntnu]{Fluids Engineering Division, Department of Energy and Process Engineering, Norwegian University of Science and Technology, N-7491 Trondheim, Norway}

\begin{abstract}
  A new mixing layer can be generated if the rotation of either of the two cylinders in a Taylor--Couette apparatus varies discontinuously along the symmetry axis. The mixing zone between the two resulting co-current streams gives rise to radial vorticity in addition to the primary axial vorticity. An analytic solution for the azimuthal velocity has been derived from which we show that the width of the mixing zone varies only with radial position.
\end{abstract}


\end{frontmatter}

\section{Introduction}

A Taylor--Couette (TC) apparatus is a popular device in which the flow of a viscous fluid between two concentric circular cylinders is driven solely by the rotation of one or both of the cylinders. The TC configuration was probably first used to determine the viscosity of water by measuring the torque on the fixed cylinder \cite{mallock1888, couette1890}. TC-apparati are still in daily use in viscometry. According to his expectations Couette found that the measured torque increased linearly with the angular velocity at low rotation rates \cite{couette1890}. The deviation observed at somewhat higher rotation rates was ascribed to a transition to turbulence. Taylor \cite{taylor23}, however, suggested that the unidirectional Couette flow became linearly unstable and an array of toroidal vortices evolved. The existence of such pairs of counter-rotating vortices, now known as Taylor vortices, was also verified experimentally by Taylor \cite{taylor23}. At even higher rotation rates, a wealth of different flow regimes may occur, see e.g.\ Refs.~\cite{andereck86,dubrulle07}.

The geometrical simplicity of the TC configuration in combination with the rich physics of the resulting flow field make it popular among applied mathematicians and physicists to explore hydrodynamic stability, transition, and turbulence.  A TC-apparatus offers an abundance of flow features in a small and closed system which makes it particularly appealing to experimentalists. The experimental investigation of migration of spherical particles suspended in a Newtonian fluid by Tetlow et al.~\cite{tetlow98} may serve as an example of a particular practical application.

In a recent study, Sprague et al.~\cite{sprague08} manipulated the inner cylinder in a TC-apparatus with a view to tailor-make the Taylor vortices. Two different concepts were explored by means of spectral-element simulations. In the type A configuration the radius of the inner cylinder varied discontinuously whereas in the type B configuration the angular velocity of the inner cylinder exhibited a discontinuous change. In this way adjacent regions of stable and unstable flow (with respect to the formation of Taylor vortices) could occur, or Taylor vortices of different wavelength on the two sides of the discontinuity could form. A related experiment with continuously changing cylinder profiles followed \cite{sprague09}.

In the present paper we are concerned only with the viscous flow which arises at relatively low rotation rates, i.e. before any toroidal Taylor vortices develop. Nevertheless, the flow will be referred to as a Taylor-Couette flow in order to distinguish it from the plane Couette flow. We consider a configuration in which the rotating cylinder is divided in two parts which rotate steadily with different angular velocities. The differently rotating parts of the cylinder give rise to adjacent flow regions with different azimuthal velocities. An explicit analytic solution will be derived for the steady flow in the mixing zone between the two co-current streams and the axial width of the mixing zone will be determined. Comparisons will also be made with a recent analytic solution for two co-current plane Couette flows \cite{narasimhamurthy11}.

\section{Mathematical formulation}\label{sec:formulation}

Let us consider the incompressible flow of a Newtonian fluid. The fluid motion in a TC-configuration is conveniently described in a cylindrical coordinate system ($r, \theta, z$) and $V_r$, $V_\theta$, and $V_z$ are the velocity components in the radial, azimuthal, and axial directions, respectively. If the fluid motion is sufficiently slow, and in the absence of end-wall effects, the velocity components in the radial and axial directions are assumed to be negligible. We are only concerned with steady-state flow. Due to the axial symmetry, the unidirectional fluid motion $V_\theta$ is independent of $\theta$. The governing equation of motion in the azimuthal direction thereby simplifies to:
\be\label{1}
  \left(\frac1{r}\pr r\pr + \pz^2 - \frac1{r^2}\right) \vt = 0.
\ee
Even though the velocity component in the radial direction $V_r$ is assumed to be negligibly small, the radial component of the Navier-Stokes equations becomes:
\be\label{2}
  \frac1{r}\vt^2 -\frac1{\rho}\pr P= 0
\ee
where $\rho$ is the fluid density. This is simply a balance between the pressure force set up in the radial direction and the centrifugal force associated the azimuthal fluid motion. The pressure field does not affect the fluid motion and $P$ can therefore be obtained from Eq.~(\ref{2}) as soon as the azimuthal velocity field $V_\theta$ has been determined by integration of Eq.~(\ref{1}). 

The fluid motion in the annular gap between inner and outer cylinders with radii $a$ and $b$, respectively, is driven by the velocities $U_A=a\Omega_A$ and $U_B=b\Omega_B$ of the inner and outer cylinder surfaces where $\Omega_A$ and $\Omega_B$ are the corresponding angular velocities. As long as the two angular velocities are constant along the axis of the TC-system, a simple closed form analytical solution can be found as a sum of a solid-body rotation $V_\theta\sim r^1$ and a potential vortex $V_\theta\sim r^{-1}$; see for instance Ref.~\cite{white06}.

In this work we are concerned with the interaction zone between two co-current Couette flows which arises if the annular velocity $\Omega$ of one or both of the cylinders changes abruptly as a function of $z$ from $\Omega^-$ to $\Omega^+$ at $z=0$. The velocity of the inner cylinder, for instance, varies discontinuously from $U_A^-=a\Omega_A^-$ at $z<0$ to $U_A^+=a\Omega_A^+$ for $z>0$. Sufficiently far away from the discontinuity we anticipate that the azimuthal velocity will depend only on the radial coordinate $r$. In the vicinity of $z=0$, on the other hand, $V_\theta=V_\theta(r,z)$, and the solution is governed by the Laplace equation (\ref{1}). In order to solve this elliptic equation, boundary conditions are required not only at the cylinder surfaces but also in the axial direction. To facilitate the derivation of an analytical solution we impose axial periodicity, i.e.,
\be
  V_\theta(r,z)=V_\theta(r,z+2L)
\ee
where $2L$ is the periodicity. At the end of the analysis the limiting case $L\to \infty$ will be considered such that the solution satisfies Neumann boundary conditions as $z\to \pm\infty$.

\section{Analytical solution}

%%%%%%%%%%%% FIGURE %%%%%%%%%%%%%
\begin{figure}[ht]
  \includegraphics[width=3.5in]{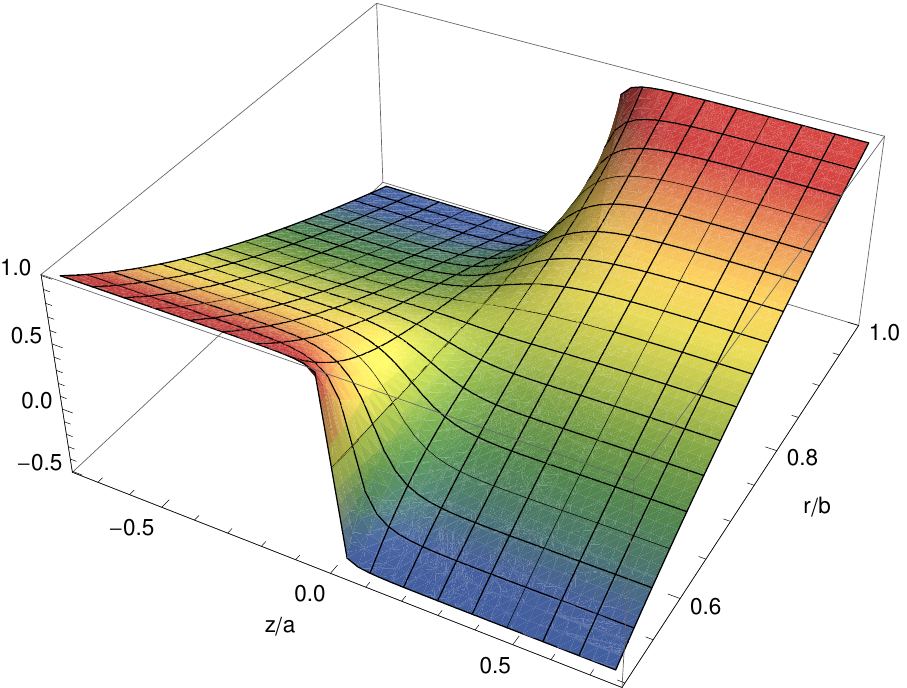} \\
  \includegraphics[width=3.5in]{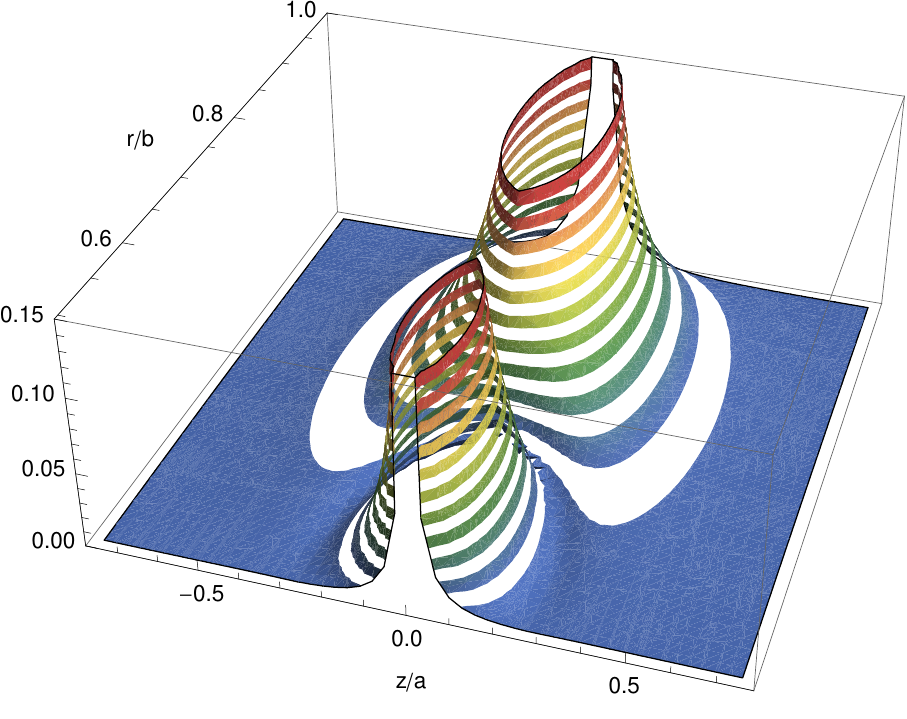} 
  \caption{Top: Asimuthal velocity $V_\theta$ as function of $r$ and $z$ for $b=2a=1$, $U_A^-=U_B^+=1$ and $U_A^+=U_B^-=-0.5$ (arbitrary units). 
  Bottom: Absolute difference between the flow in the top panel and the reference flow, Eq.~\eqref{Vref} plotted as equidistant shaded bands to illustrate mixing layer shapes with different $\delta U$ (see section \ref{sec_width}).}
  \label{fig:flow}
\end{figure}

By the standard method of separation of variables, assuming
\[
   \vt(r,z)=R(r)Z(z)
\]
gives
\begin{align*}
  Z(z) =& \left\{\begin{array}{cl} A_0 +  B_0 z, & n=0\\  A_n\cos k_n z +  B_n \sin k_n z, &n\geq 1 \end{array} \right. \\
  R(r) =& \left\{\begin{array}{cl}C_0 r+ D_0\frac1{r}, & n=0\\ C_n\Ii(k_n r)+D_n\Ki(k_n r), &n\geq 1 \end{array} \right. \\
\end{align*}
where 
\be
  k_n = \frac{n\pi}{L}, ~~ n\in \mathbb{N}_0
\ee
and $A_n, B_n, C_n, D_n$ are undetermined constants and $\mathbb{N}_0$ is the set of positive integers, including zero.

We may use standard theory of Fourier series to determine the coefficients $A_n$ and $B_n$ to satisfy the boundary conditions along the split cylinder boundaries. Combined with the remaining boundary conditions one readily obtains the analytical solution as a Fourier series:
\begin{align}
  \vt =& \langle U_A\rangle\frac{b/r-r/b}{b/a-a/b} + \langle U_B\rangle\frac{r/a-a/r}{b/a-a/b}\notag \\
  &+ \sum_{n=1}^\infty \left\{ \Delta U_A\frac{\Ki(\kr)\Ii(\kb)-\Ki(\kb)\Ii(\kr)}{\Ki(\ka)\Ii(\kb)-\Ki(\kb)\Ii(\ka)}\right.\notag \\
  &+\left.\Delta U_B\frac{\Ki(\ka)\Ii(\kr)-\Ki(\kr)\Ii(\ka)}{\Ki(\ka)\Ii(\kb)-\Ki(\kb)\Ii(\ka)}\right\}\notag\\
  &\times\frac{[1-(-1)^n]}{\pi n}\sin(\kz)\label{Fourier}
\end{align}
where
\begin{align}
  \langle{U}_{A,B}\rangle=&\half(U_{A,B}^++U_{A,B}^-)\\
  \Delta U_{A,B}=&U_{A,B}^+-U_{A,B}^-.
\end{align}
$I_1$ and $K_1$ are modified Bessel functions of order 1. One sees immediately that the first two terms (which come from the $n=0$ term) correspond to the case of normal TC flow (without cylinder splitting) in which the inner cylinder rotates at angular frequency $\langle U_A\rangle/a$ and the outer with $\langle U_B\rangle/b$. The terms proportional to $\Delta U_{A,B}$ then contain the breaking of translational symmetry along the $z$ axis, and we refer to these as the asymmetry terms.

In the special cases where either just the inner cylinder or just the outer cylinder is split, we get velocity profiles depending on a single parameter, generalising that considered in Ref.~\cite{narasimhamurthy11}, respectively,
\bs\label{FourierInnerOuter}
\begin{align}
    \frac{\vt}{a\langle \Omega_A \rangle} =& \frac{b/r-r/b}{b/a-a/b}+ \frac{\Lambda_A}{\pi}\sum_{n=1}^\infty \frac{[1-(-1)^n]}{n}\sin(\kz)\notag\\
  &\times\frac{\Ki(\kr)\Ii(\kb)-\Ki(\kb)\Ii(\kr)}{\Ki(\ka)\Ii(\kb)-\Ki(\kb)\Ii(\ka)},\label{FourierInner}
\end{align}
and 
\begin{align}
  \frac{\vt}{b\langle \Omega_B \rangle} =&\frac{r/a-a/r}{b/a-a/b}+ \frac{\Lambda_B}{\pi}\sum_{n=1}^\infty \frac{[1-(-1)^n]}{n}\sin(\kz)\notag\\
  &\times\frac{\Ki(\ka)\Ii(\kr)-\Ki(\kr)\Ii(\ka)}{\Ki(\ka)\Ii(\kb)-\Ki(\kb)\Ii(\ka)},\label{FourierOuter}
\end{align}
\es
where now $\langle \Omega_{A,B}\rangle = \half (\Omega_{A,B}^++\Omega_{A,B}^-)$ and $\Lambda_{A,B}=(\Omega_{A,B}^+-\Omega_{A,B}^-)/\langle \Omega_{A,B}\rangle$. The full solution \eqref{Fourier} may be seen to simply be the sum of \eqref{FourierInner} and \eqref{FourierOuter}, which in fact follows from the linearity of the governing equation.

In the limit $L\to \infty$ the sum over $n$ becomes an integral, and the case of Neumann boundary conditions at $z\to\pm\infty$ is obtained by means of e.g.\ the Euler--Maclaurin formula (e.g.\ Ref.~\cite{Abramowitz64}, \S 23.1) as
\begin{align}
  \vt =& \langle{U}_A\rangle\frac{b/r-r/b}{b/a-a/b} + \langle {U}_B\rangle\frac{r/a-a/r}{b/a-a/b}+ \frac{1}{\pi}\int_0^\infty \frac{\rmd\xi}{\xi}\notag \\
  &\sin(\xi z)\left\{\Delta U_A\frac{\Ki(\xi r)\Ii(\xi b)-\Ki(\xi b)\Ii(\xi r)}{\Ki(\xi a)\Ii(\xi b)-\Ki(\xi b)\Ii(\xi a)}\notag \right.\\
  &+\left.\Delta U_B\frac{\Ki(\xi a)\Ii(\xi r)-\Ki(\xi r)\Ii(\xi a)}{\Ki(\xi a)\Ii(\xi b)-\Ki(\xi b)\Ii(\xi a)}\right\},\label{IntegralResult}
\end{align}
and the special cases of either just inner or just outer cylinder split, respectively,
\bs
\begin{align}
  \frac{\vt}{a\langle \Omega \rangle} =& \frac{b/r-r/b}{b/a-a/b}+ \frac{\Lambda}{\pi}\int_0^\infty \frac{\rmd\xi}{\xi}\sin(\xi z)\notag\\
  &\times\frac{\Ki(\xi r)\Ii(\xi b)-\Ki(\xi b)\Ii(\xi r)}{\Ki(\xi a)\Ii(\xi b)-\Ki(\xi b)\Ii(\xi a)};
\end{align}
and
\begin{align}
  \frac{\vt}{b\langle \Omega \rangle} =& \frac{r/a-a/r}{b/a-a/b}+ \frac{\Lambda}{\pi}\int_0^\infty \frac{\rmd\xi}{\xi}\sin(\xi z)\notag\\
  &\times\frac{\Ki(\xi a)\Ii(\xi r)-\Ki(\xi r)\Ii(\xi a)}{\Ki(\xi a)\Ii(\xi b)-\Ki(\xi b)\Ii(\xi a)}.
\end{align}
\es

For reference in the following we will use as reference flow the (unphysical) velocity profile made up of two isolated creeping Taylor-Couette flows which meet discontinuously at $z=0$:
\bs\label{Vref}
\begin{align}
  V_\theta^\text{ref}=&\frac{b/r-r/b}{b/a-a/b}[U_A^+\Theta(z)+U_A^-\Theta(-z)]\notag \\
  &+\frac{r/a-a/r}{b/a-a/b}[U_B^+\Theta(z)+U_B^-\Theta(-z)]\\
  =&\frac{b/r-r/b}{b/a-a/b}[\langle U_A \rangle + \half\sg(z)\Delta U_A]\notag \\
  &+\frac{r/a-a/r}{b/a-a/b}[\langle U_B \rangle + \half\sg(z)\Delta U_B]
\end{align}
\es
where $\Theta(z)$ is the unit step function and $\sg(z)$ is the signum function.
The latter form is particularly useful in the following for comparison with Eq.~\eqref{IntegralResult}.

An example flow is shown in figure \ref{fig:flow}. We observe how a mixing layer forms between the two Taylor-Couette flows whose width vanishes at either wall discontinuity. 

%%%%%%%%%%%%%%%%%%%%%%%%%%%%%%%%%%%%%%%%%%%%%%
%%%%%%%%%%%%%%%%%%%%%%%%%%%%%%%%%%%%%%%%%%%%%%
%%%%%%%%%%%%%%%%%%%%%%%%%%%%%%%%%%%%%%%%%%%%%%
\subsection{The limit of plane co-current Couette flows}

As a test of our solutions, the expressions obtained in Ref.~\cite{narasimhamurthy11} for the case of a mixing layer between two plane Couette flows should be regained in the limit where 
\[
  b-a \ll a,b,r.
\]
The large sum in Eq.~(\ref{Fourier}) receives its main contributions from $n$ such that $k_n \lesssim 1/(b-a)$, since the large fraction decays exponentially for larger values of $k_n$. Then $\ka,\kr,\kb\gg 1$ and we may use (e.g.\ Ref.~\cite{Abramowitz64} \S9.7)
\[
  \Ii(z)\sim \frac{e^z}{\sqrt{2\pi z}};~~ \Ki(z)\sim \sqrt{\frac{\pi}{2z}}e^{-z},
\]
so that, with $\eta=r-a$,
\begin{align*}
  \Ki(\ka)\Ii(\kr)\sim& \frac{e^{-k_n\eta}}{2k_n\sqrt{ar}}, \\
  \Ki(\kr)\Ii(\ka)\sim& \frac{e^{k_n\eta}}{2k_n\sqrt{ar}},
\end{align*}
etc., to yield
\bs
\begin{align*}
  \frac{\Ki(\ka)\Ii(\kr)-\Ki(\kr)\Ii(\ka)}{\Ki(\ka)\Ii(\kb)-\Ki(\kb)\Ii(\ka)}\sim& \frac{\sinh k_n\eta}{\sinh 2 k_n h},\\
  \frac{\Ki(\kr)\Ii(\kb)-\Ki(\kb)\Ii(\kr)}{\Ki(\ka)\Ii(\kb)-\Ki(\kb)\Ii(\ka)}\sim& \frac{\sinh k_n(2h-\eta)}{\sinh 2 k_n h},
\end{align*}
\es
where we have used that $\sqrt{b/r}\approx \sqrt{r/a}\approx 1$. Noting additionally that
\[
  \frac{r/a-a/r}{b/a-a/b}\sim \frac{\eta}{2h}; ~~ \frac{b/r-r/b}{b/a-a/b}\sim \frac{2h-\eta}{2h}
\]
we obtain the planar result with periodic boundary conditions in the spanwise direction
\begin{align}
  V_\theta&\to U = \frac{2h-\hat z}{2h}\langle {U}_A\rangle +\frac{\hat z}{2h}\langle {U}_B\rangle\notag \\
  &+ \frac{2\Delta U_A}{\pi} 
  \sum_{n=0}^\infty\frac{\sinh\frac{(2n+1)\pi (2h-\hat z)}{L}}{\sinh\frac{(2n+1)2\pi h}{L}}\frac{\sin\frac{(2n+1)\pi \hat y}{L}}{2n+1}\notag\\
  &+ \frac{2\Delta U_B}{\pi} 
  \sum_{n=0}^\infty\frac{\sinh\frac{(2n+1)\pi \hat z}{L}}{\sinh\frac{(2n+1)2\pi h}{L}}\frac{\sin\frac{(2n+1)\pi y}{L}}{2n+1},\label{planeCouette}
\end{align}
where, for comparison with Ref.~\cite{narasimhamurthy11}, we define $\hat y=z$ and $2h=b-a$ and $\hat z=r-a$.
In the special case $\langle {U}_A\rangle =\Delta U_A=0$ this is exactly the result reported in Ref.~\cite{narasimhamurthy11}. 

In the case $L\to \infty$ one obtains the analogous Neumann boundary condition result\cite{narasimhamurthy11}
\begin{align}\label{planeCouette}
  U&(y,z) =\langle {U}_A\rangle \frac{2h-\hat z}{2h}+\langle {U}_B\rangle\frac{\hat z}{2h}  \notag \\
  &+\frac{\Delta U_A}{\pi}\arctan\left[\tanh\left(\frac{\pi \hat y}{4h}\right) \tan \left(\frac{\pi \hat (2h-\hat z)}{4h}\right)\right]  \notag \\
  &+\frac{\Delta U_B}{\pi}\arctan\left[\tanh\left(\frac{\pi \hat y}{4h}\right) \tan \left(\frac{\pi \hat z}{4h}\right)\right].
\end{align}

%%%%%%%%%%%%%%%%%%%%%%%%%%%%%%%%%%%%%%%%%%%%%%
%%%%%%%%%%%%%% S E C T I O N %%%%%%%%%%%%%%%%%
%%%%%%%%%%%%%%%%%%%%%%%%%%%%%%%%%%%%%%%%%%%%%%
\subsection{Mixing layer width}\label{sec_width}

An approximate analytical expression for the spanwise width of the mixing-layer will now be derived, similar to the analysis in Ref.~\cite{narasimhamurthy11}. Let us define the mixing-layer so that it covers the region within which the flow $V_\theta(r,z)$ in Eq.~(\ref{IntegralResult}) differs by more than some small measure $\delta U$ from the discontinuous (and unphysical) flow profile in Eq.~(\ref{Vref}) made up of the two isolated Taylor-Couette flows jumping sharply at $z=0$. 
For numerical purposes we choose $\delta U=\delta \,\mathrm{Max}(|\Delta U_A|,|\Delta U_B|)$ with $\delta$ a small parameter which we take to be $0.01$ in the numerical examples. The edges of the mixing layer, which we call $z_\text{mix}$, then solve the equation
\begin{align}
  \delta U&=\left|\frac1{\pi}\int_0^\infty\frac{\rmd \xi}{\xi}\Bigl\{\cdots\Bigr\}\sin\xi z\right.\notag\\
  &\left.-\half\sg(z)\frac{\Delta U_A(\kappa-\kappa^{-1})+\Delta U_B(\eta-\eta^{-1})}{\lambda-\lambda^{-1}}\right|
  \label{mixingEq}
\end{align}
where $\sg$ is the signum function, we make the shorthand definitions
\be
  \lambda=\frac{b}{a}, ~~ \eta=\frac{r}{a}, ~~ \kappa=\frac{b}{r}
\ee
and $\{\cdots\}$ is the expression in curly braces in Eq.~\eqref{IntegralResult}. One may show that the mixing layer is always symmetrical about $z=0$.

Noting that the integrand is an even function of $\xi$ we may expand the integration range to cover the whole real $\xi$-axis and close the contour in a large semicircle in the lower half-plane, noting that $\sin \xi z=-\sg(z)\mathrm{Im}\{\exp(-\rmi \xi|z|)\}$:
\begin{align}
  \frac1{\pi}\int_0^\infty&\frac{\rmd \xi}{\xi}\{\cdots\}\sin\xi z %\notag \\
  =  -\frac{\sg(z)}{2\pi}\mathrm{Im} \int_{-\infty}^\infty\frac{\rmd \xi}{\xi}\{\cdots\}e^{-\rmi \xi|z|} \notag \\
  =&\frac{\sg(z)}{2\pi}\mathrm{Im} \oint_\Gamma\frac{\rmd \xi}{\xi}\{\cdots\}e^{-\rmi \xi|z|}\notag \\
  =&\sg(z)\mathrm{Re} \sum_{j=0}^\infty \Bigl.^\prime \underset{\xi=\xi_j}{\rm Res} \frac{1}{\xi}\{\cdots\}e^{-\rmi \xi|z|}\label{oint}
\end{align}
where $\Gamma$ is the integration path previously described, $j$ sums over the poles $\xi_j$ of the integrand, and the prime on the summation mark signifies that the pole at $\xi=0$ is counted with half weight. These poles lie at $\xi=0$ and along the imaginary $\xi$-axis and the ones along the negative imaginary axis we denote $\xi_j=-\rmi y_j$. Calculating the residue at $\xi_0=0$ we find that it exactly cancels the second term on the right-hand side of Eq~\eqref{mixingEq}. 

For the remainder of the sum we make the approximation of keeping only the first term, corresponding to $\zeta_1=ay_1$ which is the smallest of the values $\zeta_n$ that solve
\be\label{zetaEq}
  \frac{Y_1(\zeta_j)}{Y_1(\lambda \zeta_j)}=\frac{J_1(\zeta_j)}{J_1(\lambda \zeta_j)}
\ee
where $J_n$ and $Y_n$ are Bessel functions. We then obtain the solution
\begin{align}\label{zmixFirst}
  |z_\text{mix}|\approx&-\frac{a}{\zeta_1}\log\frac{\delta U}{F(\zeta_1)}
\end{align}
where
\begin{align}
  F(\zeta) =& \frac{2}{\lambda \zeta^2}\frac{\Delta U_A p_1(\eta \zeta,\lambda \zeta)+\Delta U_B p_1(\zeta,\eta \zeta)}{p_2(\zeta,\lambda \zeta)-p_0(\zeta,\lambda \zeta)},\\
  p_n(x_1,x_2)=&J_n(x_1)Y_n(x_2)-J_n(x_2)Y_n(x_1),
\end{align}
and we have used (Ref.~\cite{Abramowitz64} \S9.1)
\[
  \frac{\rmd }{\rmd x}p_1(ax,bx)=-\half abx[p_2(ax,bx)-p_0(ax,bx)].
\]

%%%%%%%%%%%%%%%%%%%%% FIGURE %%%%%%%%%%%%%%%%%%%%%%%%%%
\begin{figure*}
  \includegraphics[width=7in]{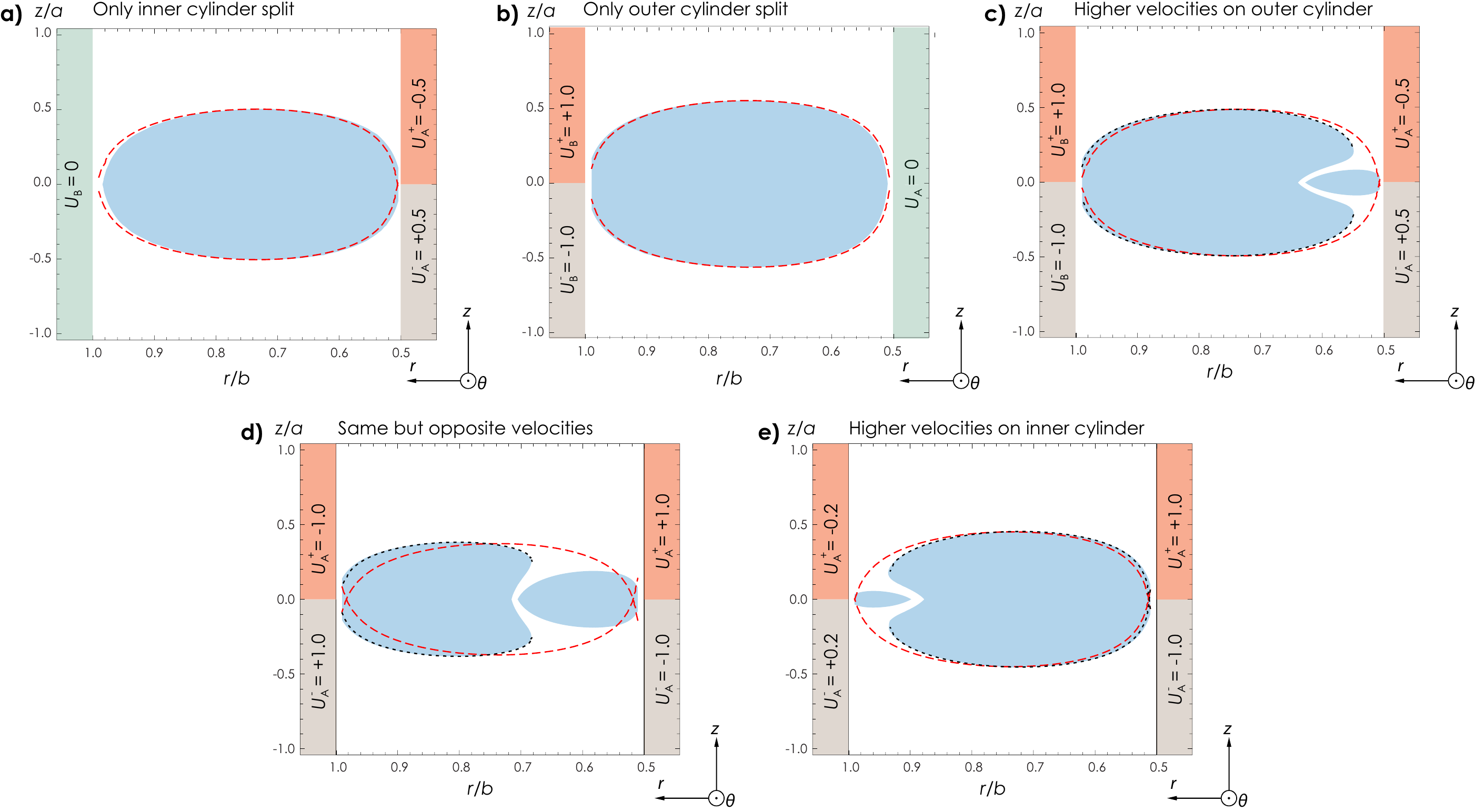}
  \caption{Mixing layer for the three cases of only inner cylinder split, only outer cylinder split, and both cylinders split. The non-split cylinder is at rest (but this makes no difference to the mixing layer width). Velocities in arbitrary units. The exact boundary layer is the shaded area and the dashed line shows the first-order analytical approximation, Eq.~\eqref{zmixFirst}. In panels c, d and e also the second order approximation Eq.~\eqref{zmixSecond} is shown as a dotted line. $\delta U = 0.01 \max_{A,B}(\Delta U)$ and $b=2a=1$ in all panels.}
  \label{fig:mixingpanels}
\end{figure*}

We can obtain a better approximation by including one further term of the infinite sum of residues and use that $\zeta_2\approx 2\zeta_1$ [for large $\zeta$ the solutions to \eqref{zetaEq} are asymptotically $\zeta_n\sim n\pi/(\lambda-1)$], which we will refer to as a second order approximation:
\be\label{zmixSecond}
    |z_\text{mix}|\approx-\frac{a}{\zeta_1}\log\frac{s_1\sqrt{F^2(\zeta_1)+4\delta U s_1F(\zeta_2)}-F(\zeta_1)}{2F(\zeta_2)}
\ee
with $s_1=\mathrm{Sg}[F(\zeta_1)]$.
This tends to the first order expression \eqref{zmixFirst} when $\delta U\to 0$.

Mixing layers for a few different cases are shown in Fig~\ref{fig:mixingpanels}. 
The shaded area represents the mixing zone calculated from the exact analytical solution in Eq.~(\ref{Fourier}), or if appropriate, from Eq.~(\ref{FourierInnerOuter}). Outside of the circumference of the shaded area, the exact solution deviates from the approximate discontinuous solution in Eq.~(\ref{Vref}) by less than 1 per cent. 
When both cylinders are split, two-lobed mixing layer shapes are observed, as shown in panels c-e in the figure. In this case the second order approximation \eqref{zmixSecond} follows the shape of the greater of the lobes with excellent precision whereas the first order approximation \eqref{zmixFirst}, although taking the shape of a single lobe, gives a reasonably good outline of the contour of the mixing layer albeit with somewhat limited accuracy when the two lobes are of similar size, as exemplified in panel d. The maximum width of the mixing layer is however well predicted by Eq.~\eqref{zmixFirst}.

%%%%%%%%%%%%%%%%%%%%%%%%%%%%%%%%%%%%%%%%%%%%%%
%%%%%%%%%%%%%% S E C T I O N %%%%%%%%%%%%%%%%%
%%%%%%%%%%%%%%%%%%%%%%%%%%%%%%%%%%%%%%%%%%%%%%
\section{Shear stress and torque on rotating cylinders}\label{sec:torque}

It is of interest to investigate how the mixing layer changes the viscous torque experienced by the rotating cylinders as compared to the standard Taylor-Couette set-up in the creeping flow regime. The shear stress acting on the cylinder surfaces at $r=a,b$ directed along the local $\theta$-axes are obtained via the rates of strain from Newton's law of viscous friction,
\be
  \tau_A = \mu \partial_r V_\theta\Bigr|_{r=a}, ~~ \tau_B =-\mu\partial_r V_\theta\Bigr|_{r=b}
\ee
with $\mu$ being the dynamic viscosity.
The torques acting on the walls in the standard Taylor-Couette set-up are easily calculated, and generalizing slightly to the reference flow in Eq.~\eqref{Vref}, the $z$ component (being the only nonzero component) is
\bs\label{dVref}
\begin{align}
  M_A^{\pm,\text{ref}}=& 2\pi \mu a L\frac{2U_B^\pm-(\lambda+\lambda^{-1})U_A^\pm}{\lambda-\lambda^{-1}},\\
  M_B^{\pm,\text{ref}}=& 2\pi \mu b L\frac{2U_A^\pm-(\lambda+\lambda^{-1})U_B^\pm}{\lambda-\lambda^{-1}}.
\end{align}
\es
Here $2L$ is the total length of the split cylinders.

\begin{figure}[t]
  \includegraphics[width=3.5in]{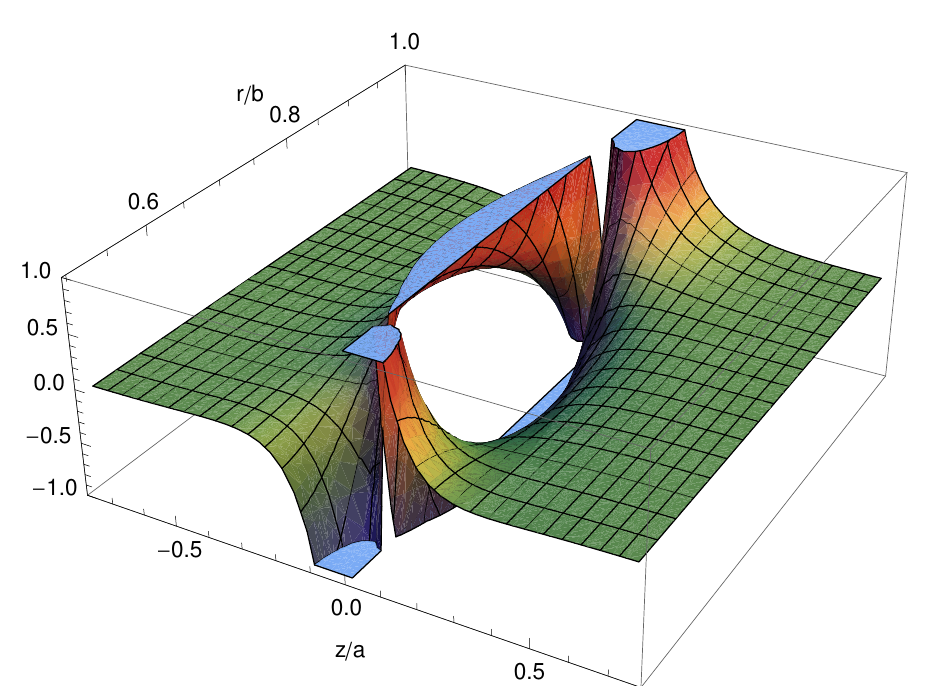}\\
  \includegraphics[width=3.5in]{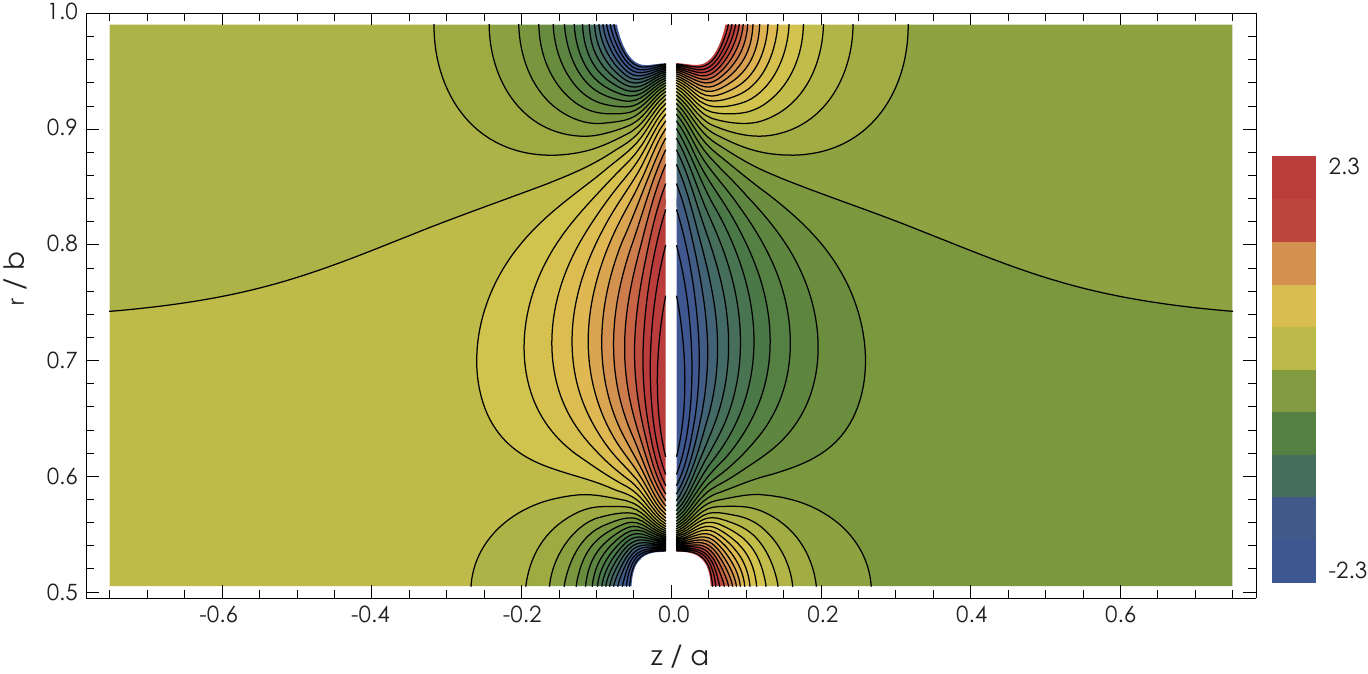}
  \caption{Radial derivative of $V_\theta$ with same parameter values as in figure \ref{fig:flow} and the contours of the same (bottom). The radial derivative has a pole singularity at the two discontinuity points.}
  \label{fig:derivative}
\end{figure}

It turns out that the additional torque on the individual halves of a split cylinder due to the presence of the mixing layer is formally infinite. Mathematically this can be seen from the fact that $\partial_r V_\theta$ at $r=a$ and $b$ has a simple pole-like singularity $\sim 1/z$ at $z=0$. An integral over the cylindrical surface area integrating to $z=0$ must thus necessarily be logarithmically divergent. The same is found to be true of the shear force in the planar case considered in Ref.~\cite{narasimhamurthy11}, Eq.~\eqref{planeCouette}. The radial derivative of $V_\theta$ is shown in Fig.~\ref{fig:derivative} for the same parameters as in Fig.~\ref{fig:flow}. Clearly, this infinity is not physical, a point to which we shall return. We will discuss the behaviour of $\partial_r V_\theta$ and the local shear stress further in the following.

An explicit expression for $\partial_r V_\theta$ is trivial to write down from Eq.~\eqref{IntegralResult}. We are interested in the difference in viscous shear resistance due to the mixing layer, compared to that of the reference flow field \eqref{Vref}, as given by Eqs.~\eqref{dVref}, and subtracting the derivative corresponding to the reference case the difference may be written
\begin{align}
  \Delta&\partial_rV_\theta=\frac{1}{\pi}\int_0^\infty \rmd\xi\left\{\Delta U_A\frac{\Ki'(\xi r)\Ii(\xi b)-\Ki(\xi b)\Ii'(\xi r)}{\Ki(\xi a)\Ii(\xi b)-\Ki(\xi b)\Ii(\xi a)}\right.\notag \\
  &+\left.\Delta U_B\frac{\Ki(\xi a)\Ii'(\xi r)-\Ki'(\xi r)\Ii(\xi a)}{\Ki(\xi a)\Ii(\xi b)-\Ki(\xi b)\Ii(\xi a)}\right\}\sin \xi z\notag \\
  &-\frac{\sg(z)}{2r}\Bigl[\frac{\Delta U_B(\eta+\eta^{-1})-\Delta U_A(\kappa+\kappa^{-1})}{\lambda-\lambda^{-1}}\Bigr].\label{Ddvdr}
\end{align}
It is convenient in the following to move the last term into the integral over $\xi$ by noting the Fourier identity
\[
  \sg(z)=\frac{2}{\pi}\int_0^\infty \frac{\rmd \xi}{\xi}\sin\xi z
\]
which is valid for any real $z$.

At the inner and outer cylinders Eq.~\eqref{Ddvdr} reads, explicitly,
\bs\label{DdvdrAB}
\begin{align}
  &\Delta\partial_rV_\theta\Bigr|_{r=a}=\frac{1}{\pi}\int_0^\infty \rmd\xi\sin \xi z\left\{\frac{\Delta U_A(\lambda+\lambda^{-1})-2\Delta U_B}{\xi a(\lambda-\lambda^{-1})}\right.\notag \\
  &\left.+\frac{\Delta U_B/\xi a+\Delta U_A[\Ki'(\xi a)\Ii(\xi b)-\Ki(\xi b)\Ii'(\xi a)]}{\Ki(\xi a)\Ii(\xi b)-\Ki(\xi b)\Ii(\xi a)}\right\} \label{DdvdrA}
\end{align}
and
\begin{align}
  &-\Delta\partial_rV_\theta\Bigr|_{r=b}=\frac{1}{\pi}\int_0^\infty \rmd\xi\sin \xi z\left\{\frac{\Delta U_B(\lambda+\lambda^{-1})-2\Delta U_A}{\xi b(\lambda-\lambda^{-1})}\right.\notag \\
  &+\left.\frac{\Delta U_A/\xi b+\Delta U_B[\Ki'(\xi b)\Ii(\xi a)-\Ki(\xi a)\Ii'(\xi b)]}{\Ki(\xi a)\Ii(\xi b)-\Ki(\xi b)\Ii(\xi a)}\right\}.\label{DdvdrB}
\end{align}
\es

We may use the path integral technique as in the previous section to write this as a sum of residues. The manipulation \eqref{oint} holds also for the integrals in \eqref{Ddvdr} and \eqref{DdvdrAB}. Calculating the residues, which are found at the same poles as in the previous section, we again find that the residue of the pole at $z=0$ is exactly cancelled between the two terms in the integrands of Eqs.~\eqref{DdvdrA} and \eqref{DdvdrB}. The derivatives may then be written as sums over the remaining residues,
\bs\label{strainratesRes}
\begin{align}
  \Delta\partial_rV_\theta\Bigr|_{r=a}=&\frac{2}{b}\sg(z)\sum_{j=1}^\infty e^{-\zeta_j |z|/a}\notag \\
  &\times\frac{\Delta U_A\zeta_jq_1(\lambda\zeta_j,\zeta_j)-\Delta U_B}{\zeta_j^2[p_2(\zeta_j,\lambda \zeta_j)-p_0(\zeta_j,\lambda \zeta_j)]}
\end{align}
and
\begin{align}
  -\Delta\partial_rV_\theta\Bigr|_{r=b}=&\frac{2}{b}\sg(z)\sum_{j=1}^\infty e^{-\zeta_j |z|/a}\notag \\
  &\times\frac{\Delta U_B\zeta_jq_1(\zeta_j,\lambda\zeta_j)-\Delta U_A/\lambda}{\zeta_j^2[p_2(\zeta_j,\lambda \zeta_j)-p_0(\zeta_j,\lambda \zeta_j)]}
\end{align}
\es
with
\be
  q_n(x,y) = J_n(x)Y_n'(y)-J_n'(y)Y_n(x).
\ee
The values $\zeta_j$ are still solutions of Eq.~\eqref{zetaEq}. When $|z|$ is of order $a$ or higher, these sums converge rapidly and are numerically much cheaper than evaluating the integrals. 

As already mentioned, the torque contributions to the half-cylinders from the singular points are formally infinite when the shear stress is integrated over the cylinder surfaces all the way to $z=0$. Let us therefore exclude a thin band of width $\Delta z$ close to the discontinuity and calculate the torques on the upper (+) and lower (-) cylinders according to
\begin{align*}
  M^+_A =& 2\pi \mu a^2 \int_{\Delta z}^\infty \rmd z\, \partial_r V_\theta\Bigr|_{r=a}; \\
  M^-_A =& 2\pi \mu a^2 \int_{-\infty}^{-\Delta z} \rmd z\, \partial_r V_\theta\Bigr|_{r=a},
\end{align*}
and similarly for the outer cylinders.

With expressions \eqref{strainratesRes} for the strain rates the corresponding torques on the upper (+) and lower (-) inner (A) and outer (B) cylinders then take the form
\bs
\begin{align}
  \frac{\Delta M_A^\pm}{\Delta M_A^{(0)}}=&\pm\frac{2}{\lambda}\sum_{j=1}^\infty \frac{[\zeta_jq_1(\lambda\zeta_j,\zeta_j)-\frac{\Delta U_B}{\Delta U_A}]e^{-\zeta_j \Delta z/a}}{\zeta_j^3[p_2(\zeta_j,\lambda \zeta_j)-p_0(\zeta_j,\lambda \zeta_j)]}
\end{align}
and
\begin{align}
    \frac{\Delta M_B^\pm}{\Delta M_B^{(0)}}=&\pm\frac{2}{\lambda}\sum_{j=1}^\infty \frac{[\zeta_jq_1(\zeta_j,\lambda\zeta_j)-\frac{\Delta U_A}{\lambda\Delta U_B}]e^{-\zeta_j \Delta z/a}}{\zeta_j^3[p_2(\zeta_j,\lambda \zeta_j)-p_0(\zeta_j,\lambda \zeta_j)]}.
\end{align}
\es
We have defined the reference torques
\be
  \Delta M_A^{(0)}=2\pi\mu a^2 \Delta U_A, ~~   \Delta M_B^{(0)}=2\pi\mu b^2 \Delta U_B.
\ee

As $\Delta z\to 0$ the sum is dominated by large values of $\zeta_j\approx \pi j/(\lambda-1)$, for which, using asymptotic expansions of Bessel functions (Ref.~\cite{Abramowitz64} \S 9.2)
\begin{align*}
  \frac{2}{\lambda}\frac{\zeta_jq_1(\zeta_j,\lambda\zeta_j)-\frac{\Delta U_B}{\Delta U_A}}{\zeta_j^3[p_2-p_0](\zeta_j,\lambda \zeta_j)}
  \sim&  -\frac{1}{j}\left[\frac1{\pi}-\frac{\sqrt{\lambda}}{2}(-1)^{j}\frac{\Delta U_B}{\Delta U_A}\right]\\
  \frac{2}{\lambda}\frac{[\zeta_jq_1(\zeta_j,\lambda\zeta_j)-\frac{\Delta U_A}{\lambda\Delta U_B}]}{\zeta_j^3[p_2-p_0](\zeta_j,\lambda \zeta_j)}
  \sim&  -\frac{1}{j}\left[\frac1{\pi}-\frac{1}{2\sqrt{\lambda}}(-1)^{j}\frac{\Delta U_A}{\Delta U_B}\right]
\end{align*}
where we noted 
\begin{align*}
  q_1(\zeta_j,\lambda\zeta_j)\sim&\, q_1(\lambda\zeta_j,\zeta_j)\sim \frac{2}{\pi\zeta_j\sqrt{\lambda}}\cos[\zeta_j(\lambda-1)]\\
  [p_2-p_0](\zeta_j,\lambda\zeta_j)\sim& -\frac{4(\lambda-1)}{\pi\zeta_j^2\lambda^{3/2}}\cos [\zeta_j(\lambda-1)]
\end{align*}
for large $\zeta_j$. We have inserted $\zeta_j\approx j\pi/(\lambda-1)$ for large $j$. The shorthand notation introduced in the denominators should cause no confusion. 
We obtain 
\bs
\begin{align}
  \frac{\Delta M_A^\pm}{\Delta M_A^{(0)}}\sim& \mp \frac1{2} \sqrt{\frac{b}{a}}\frac{\Delta U_B}{\Delta U_A}\log\frac{\pi\Delta z}{b-a}\\
  \frac{\Delta M_B^\pm}{\Delta M_A^{(0)}}\sim& \mp \frac1{2} \sqrt{\frac{a}{b}}\frac{\Delta U_A}{\Delta U_B}\log\frac{\pi\Delta z}{b-a}
\end{align}
\es
for $\Delta z\to 0$.

In the mathematical formulation of the problem in Section \ref{sec:formulation}, the angular velocity of either the inner or the outer cylinder (or both) was assumed to exhibit a discontinuous change at $z = 0$. This discontinuity gave rise to a pole-like singularity when the wall shear stresses and the accompanying torques were calculated in Section \ref{sec:torque}. In practice, however, a discontinuous change of the rotation rate is a mathematical artifact which cannot be reproduced in a laboratory TC apparatus. If the inner cylinder, say, is divided in two differently rotating halves, a tiny gap will exist between the two parts and tend to smear out what was meant as a discontinuity in the angular velocity of  the cylinder. 

%%%%%%%%%%%%%%%%%%%%%%%%%%%%%%%%%%%%%%%%%%%%%%
%%%%%%%%%%%%%% S E C T I O N %%%%%%%%%%%%%%%%%
%%%%%%%%%%%%%%%%%%%%%%%%%%%%%%%%%%%%%%%%%%%%%%
\section{Concluding remarks}

An analytic solution has been derived for the shear layer between two co-current Taylor--Couette flows. A steady-state solution which obeyed axial periodicity was obtained first and thereafter simplified to an infinitely long period. In the limit of large radii and a narrow gap, the earlier solution by Narasimhamurhty et al.\ \cite{narasimhamurthy11} for the bilateral shear layer between two parallel plane Couette flows was recovered. 

While co-current plane Couette flows are challenging to realize in a laboratory experiment, a Taylor--Couette flow apparatus can be constructed such as to allow the inner cylinder to rotate with an almost discontinuous angular velocity. This set-up will enable investigations of, for instance, particle dispersion or chemical reactions in a bilateral shear flow with shearing motion not only in the $(r, \theta)$-plane both also in the perpendicular $(\theta,z)$-plane.

%%%%%%%%%%%%%%%%%%%%%%%%%%%%%%%%%%%%%%%%%%%%%%%%%%%%%%%%%%%%%%%%%%%

\end{document}